\documentclass[12pt,aps,showpacs,notitlepage]{revtex4-1}
\usepackage{amssymb,amsmath}
\newcommand{\beq}{\begin{equation}}
\newcommand{\beqn}{\begin{eqnarray}}
\newcommand{\eeq}{\end{equation}}
\newcommand{\eeqn}{\end{eqnarray}}
\begin{document}

\title{Primordial Perturbations from Multifield Inflation with Nonminimal Couplings}
\author{David I. Kaiser}
\email{Email: dikaiser@mit.edu}
\affiliation{Center for Theoretical Physics and Department of Physics, \\
Massachusetts Institute of Technology, Cambridge, Massachusetts 02139 USA}
\author{Audrey T. Todhunter}
\email{Email: audrey.todhunter@gmail.com}
\affiliation{Institut de Th\'{e}orie des Ph\'{e}nom\`{e}nes Physiques, \'{E}cole Polytechnique F\'{e}d\'{e}rale de Lausanne, \\ CH-1015 Lausanne, Switzerland}
\date{\today}

\begin{abstract} 
Realistic models of particle physics include many scalar fields. These fields generically have nonminimal couplings to the Ricci curvature scalar, either as part of a generalized Einstein theory or as necessary counterterms for renormalization in curved background spacetimes. We develop a gauge-invariant formalism for calculating primordial perturbations in models with multiple nonminimally coupled fields. We work in the Jordan frame (in which the nonminimal couplings remain explicit) and identify two distinct sources of entropy perturbations for such models. One set of entropy perturbations arises from interactions among the multiple fields. The second set arises from the presence of nonminimal couplings. Neither of these varieties of entropy perturbations will necessarily be suppressed in the long-wavelength limit, and hence they can amplify the curvature perturbation, $\zeta$, even for modes that have crossed outside the Hubble radius. Models that overproduce long-wavelength entropy perturbations endanger the close fit between predicted inflationary spectra and empirical observations.
\end{abstract}
\pacs{04.62+v; 98.80.Cq. Preprint MIT-CTP 4143  \\ Published as Phys. Rev. D81, 124037 (2010).}
\maketitle

\section{Introduction} 

High-precision measurements of the Cosmic Microwave Background (CMB) radiation provide some of the most stringent tests of inflationary cosmology \cite{Komatsu,GuthKaiser}. As is well known, inflationary models make specific predictions for the spectrum of primordial perturbations. The gauge-invariant formalism for performing those calculations, based on coupled metric perturbations and quantum field fluctuations, has reached a high state of maturity. (For reviews, see \cite{MFB,RHB2003,BTW,LiddleLythBook,MalikWands}.) 

An important feature of the perturbation spectra concerns the balance between adiabatic perturbations and isocurvature (or entropy) perturbations. Models that overproduce entropy perturbations can result in significant differences between predicted inflationary spectra and empirical observations \cite{GWBM,BassettTsujikawa,BTW}. In particular, entropy perturbations can amplify the curvature perturbation on the longest (cosmologically relevant) length scales, even after those modes have crossed outside the Hubble radius. Single-field models generically predict little to no entropy perturbations on the longest length scales. But models that involve multiple interacting scalar fields do, in general, produce long-wavelength entropy perturbations, which can threaten the conservation of the curvature perturbation and hence the matching of inflationary predictions to observations \cite{GWBM,WandsReview}.

The authors of \cite{RFB2002} identified a second source of entropy perturbations: scalar fields with noncanonical kinetic terms will induce entropy perturbations distinct from the usual source that stems from fields' interactions with each other. Building on that important observation, recent work \cite{Spinflation,Langlois} has considered perturbations in generalized multifield inflationary models, in which arbitrarily many scalar fields possess noncanonical kinetic terms. The analyses in \cite{RFB2002,Spinflation,Langlois} were conducted in the Einstein frame, in which all fields have minimal couplings to the Ricci curvature scalar, $R$.

Noncanonical kinetic terms are often associated with exotic forms of matter (such as axions and moduli fields from string theory), if not outright pathologies. Scalar fields with the ``wrong" sign of the kinetic term, for example, can signal a tachyonic instability in a model. On the other hand, noncanonical kinetic terms also appear upon making a conformal transformation to the Einstein frame and rescaling any fields which, in the Jordan frame, possessed nonminimal couplings to $R$ --- even if those fields had canonical kinetic terms in the Jordan frame. As is well known, nonminimal couplings are generic for scalar fields. They arise from a variety of model-building efforts (in supergravity, string theory, and more) \cite{Fujii,Faraoni2004}. They are also required as counterterms when considering renormalization of scalar fields in curved background spacetimes \cite{BirrellDavies,Buchbinder}. Indeed, in many models the nonminimal coupling strength, $\xi$, grows without bound under renormalization-group flow \cite{Buchbinder}.

As has recently been shown \cite{DKconftrans}, for models that incorporate multiple nonminimally coupled scalar fields, there does not exist any combination of conformal transformation and field rescalings that can bring both the gravitational portion of the action and the fields' kinetic terms into canonical form. Thus one may work in a frame that incorporates canonical Einstein gravity but necessarily includes noncanonical kinetic terms for the scalar fields; or one may work in a frame in which the nonminimal couplings remain explicit but the scalar fields have canonical kinetic terms. 

Here we revisit the calculation of \cite{Spinflation,Langlois}, performing the entire calculation in the Jordan frame. By working in the Jordan frame, the second source of entropy perturbations described in \cite{RFB2002,Spinflation,Langlois} appears rather straightforward: it arises from the presence of the nonminimal couplings, and persists even for scalar fields that retain canonical kinetic terms (in the Jordan frame). That is, models that have a perfectly mundane matter sector -- each scalar field with a canonical kinetic term, and each with a simple nonminimal coupling (as required for renormalization) --- include the second source of entropy perturbations identified in \cite{RFB2002,Spinflation,Langlois}. In the process, we build on previous work that extended the gauge-invariant perturbation formalism to generalized Einstein theories \cite{Hwang,MFB,HiraiMaeda,NewTwist,NMCPerts} to present what is (to the best of our knowledge) the first calculation of inflationary perturbations for models with multiple nonminimally coupled fields, conducted in the Jordan frame in which the nonminimal couplings remain explicit. (See \cite{Faraoni2001,Faraoni2004} for reviews of the single-field, nonminimally coupled case.)

We perform the calculation in the Jordan frame because we believe other aspects of inflationary model-building may be tackled more conveniently in the Jordan frame. In addition to the fields' canonical kinetic terms, interactions among the scalar fields may be readily analyzed without the appearance of new (and often non-renormalizable) interactions among the transformed fields, which necessarily appear in multifield models upon making use of a conformal transformation \cite{BurgessHertzberg,DKconftrans}. 

The formalism developed here should be helpful for calculating primordial perturbation spectra for all manner of models of recent interest, whether inspired by string cosmology or not. Indeed, because nonminimal couplings are generic for scalar fields in curved spacetime --- and because realistic models of particle physics (including the Standard Model and its various generalizations) contain many scalar fields that could play important roles in the early universe \cite{LythRiottoMazumdar} --- it is essential to have a robust, gauge-invariant formalism that can accommodate multiple nonminimally coupled fields. We have in mind, for example, the recent model of ``Higgs inflation," in which the Higgs sector of the electroweak Standard Model drives a phase of early-universe inflation, thanks to a significant nonminimal coupling, $\xi \sim 10^4$ \cite{ShapBez}. In renormalizable gauges appropriate to the high-energy inflationary regime, the Standard Model Higgs sector includes four scalar fields rather than just one: the (real) Higgs scalar field plus three Goldstone fields \cite{WeinbergQFT,BurgessHertzberg,DKconftrans}. ``Higgs inflation" thus involves four nonminimally coupled scalar fields, each of which possesses a canonical kinetic term in the Jordan frame. The formalism developed here will allow one to analyze perturbation spectra for models like these.

The remainder of the paper is organized as follows. In Section II we derive the equations governing linearized metric perturbations for the general case of multiple nonminimally coupled scalar fields. In Section III we examine the curvature perturbation on uniform-density hypersurfaces for such models, $\zeta$, to highlight under what conditions $\zeta$ may vary considerably between the time a mode first crosses outside the Hubble radius during inflation and the time that mode re-enters the Hubble radius at the epoch of CMB last scattering. Concluding remarks follow in Section IV.

\section{Background and Perturbations}

We work in $(3 + 1)$ spacetime dimensions, with metric signature $(-, +,+,+)$. We consider models involving $N$ scalar fields, $\phi^I$, each of which is nonminimally coupled to the Ricci curvature scalar. The action is given by
\beq
S = \int d^4 x \sqrt{-g} \left[ f (\phi^I) R - \frac{1}{2} {\cal G}_{IJ} g^{\mu\nu} \phi^I_{; \mu} \phi^J_{; \nu} - V (\phi^I) \right] .
\label{action}
\eeq
We denote covariant derivatives (with respect to the spacetime metric, $g_{\mu\nu}$) with semicolons, and use the Einstein summation convention both for repeated spacetime indices ($\mu, \nu$) and field-space indices ($I, J$). Since our goal is to demonstrate the relationship between entropy perturbations and nonminimal couplings, we will restrict attention to trivial (Euclidean) field spaces, with ${\cal G}_{IJ} = \delta_{IJ}$. Hence every scalar field in the models we consider has a canonical kinetic term in the action. More complicated field-space metrics may be treated by the methods developed in \cite{Spinflation,Langlois}. 

Varying the action with respect to $g_{\mu\nu}$ yields the Einstein field equations
\beq
G_{\mu\nu} = R_{\mu\nu} - \frac{1}{2} g_{\mu\nu} R = \frac{1}{M_{\rm pl}^2} T_{\mu\nu} , 
\label{Rmn}
\eeq
where $M_{\rm pl}$ is the reduced Planck mass,
\beq
M_{\rm pl} \equiv \frac{1}{\sqrt{8\pi G}} = 2.43 \times 10^{18} \> {\rm GeV} .
\label{Planckmass}
\eeq
The energy-momentum tensor takes the form
\beq
T_{\mu\nu} = \frac{M_{\rm pl}^2}{2 f (\phi^I)} \left[  \delta_{IJ} \phi^I_{; \mu} \phi^J_{; \nu} - g_{\mu\nu} \left( \frac{1}{2} \delta_{KL} g^{\lambda \sigma} \phi^K_{; \lambda} \phi^L_{; \sigma} + V (\phi^I) \right)  + 2 f(\phi^I)_{; \mu; \nu} - 2 g_{\mu\nu} \Box f (\phi^I) \right] .
\label{Tmn}
\eeq
The coefficient $M_{\rm pl}^2 / (2f)$ on the righthand side of Eq. (\ref{Tmn}) is necessary in order for $T_{\mu\nu}$ to be conserved, $T^{\mu\nu}_{\>\>\>\>; \nu} = 0$.

Varying the action with respect to each field $\phi^I$ yields the equations of motion
\beq
\Box \phi^I - V_I  + f_I  R = 0 ,
\label{eomJ}
\eeq
where $\Box \phi^I = g^{\mu\nu} \phi^I_{; \mu ; \nu}$, $V_I = \partial V / \partial \phi^I$, and $f_I = \partial f / \partial \phi^I$. Because both $V$ and $f$ depend on multiple scalar fields, $\phi^I, \phi^J, ... , \phi^N$, the fields' equations of motion will, in general, become coupled.

We may next consider linearized perturbations around a spatially flat Friedmann-Robertson-Walker metric in $(3 + 1)$ spacetime dimensions. The scalar degrees of freedom of the perturbed line element may be written
\beq
ds^2 = - (1 + 2 A) dt^2 + 2a (\partial_i B ) dx^i dt + a^2 \left[ (1 - 2\psi) \delta_{ij} + 2 \partial_i \partial_j E \right] dx^i dx^j ,
\label{ds2}
\eeq
where Latin indices run over spatial coordinates, $i, j = 1, 2, 3$, and $a(t)$ is the scale factor. The well-known gauge-invariant Bardeen potentials are defined as \cite{MFB,BTW}
\beq
\begin{split}
\Phi &\equiv A - \frac{d}{dt} \left[ a^2 \left( \dot{E} - \frac{B}{a} \right) \right] , \\
\Psi &\equiv \psi + a^2 H \left( \dot{E} - \frac{B}{a} \right) ,
\end{split}
\label{Bardeen}
\eeq
where overdots denote derivatives with respect to cosmic time, $t$, and $H = \dot{a} / a$ is the Hubble parameter. In the longitudinal gauge, which corresponds to setting $E = B = 0$, the perturbed line element reduces to
\beq
ds^2 = - (1 + 2 \Phi) dt^2 + a^2 (1 - 2 \Psi) \delta_{ij} dx^i dx^j .
\label{dslong}
\eeq
We may also separate each scalar field into a spatially homogenous background and a fluctuation:
\beq
\phi^I (x^\mu) = \varphi^I (t) + \delta \phi^I (x^\mu) .
\label{fieldfluctuations}
\eeq
In the spacetime metric associated with Eq. (\ref{dslong}), the equations of motion, Eq. (\ref{eomJ}), separate into background and first-order expressions:
\beq
\begin{split}
\ddot{\varphi}^I &+ 3 H \dot{\varphi}^I + V_I (\varphi^I) - f_I (\varphi^I) R  = 0 , \\
\delta \ddot{\phi}^I &+ 3 H \delta \dot{\phi}^I -  \frac{1}{a^2} \nabla^2 \delta \phi^I + \left( V_{IJ} - f_{IJ} R \right) \delta \phi^J \\
&\quad\quad\quad = - 2 V_I \Phi + \dot{\varphi}^I \left( \dot{\Phi} + 3 \dot{\Psi} \right) + f_I \left( 2 R \Phi +  \delta R  \right),
\end{split}
\label{eomvarphideltaphi}
\eeq
where $\nabla^2 = \partial^i \partial_i$ is the spatial Laplacian in comoving coordinates, and
\beq
\begin{split}
R &= 6 \left( \dot{H} + 2 H^2 \right) , \\
\delta R &= - 6 \ddot{\Psi} - 6H \left( \dot{\Phi} + 4 \dot{\Psi} \right) - 12 \left( \dot{H} + 2 H^2 \right) \Phi -  \frac{2}{a^2} \nabla^2 \left( \Phi - 2 \Psi \right) .
\end{split}
\label{RdeltaR}
\eeq

We expand the Einstein field equations to first order in $\Phi$, $\Psi$, and $\delta \phi^I$,
\beq
G_{\mu\nu}^{(0)} + \delta G_{\mu\nu} = \frac{1}{M_{\rm pl}^2} \left[ T_{\mu\nu}^{(0)} + \delta T_{\mu\nu} \right] ,
\label{Einstein}
\eeq
with background quantities labeled by $(0)$. It will prove convenient to write the energy-momentum tensor in terms of fluid quantities first, and later use Eq. (\ref{Tmn}) to relate quantities such as the energy density, $\rho$, and the pressure, $p$, to the fields and their fluctuations. We follow the convention of labeling fluid quantities and their perturbations in terms of the mixed-index energy-momentum tensor:
\beq
\begin{split}
T^0_{\>\>0} &= - ( \rho + \delta \rho ) , \\
T^0_{\>\>i} &= \partial_i \delta q , \\
T^i_{\> \>j} &= \delta^i_{\> j} (p + \delta p) + \Pi^i_{\>\> j} ,
\end{split}
\label{fluid}
\eeq
where $\delta \rho$ is the density perturbation, $\delta q$ is the momentum flow, $\delta p$ is the isotropic pressure perturbation, and $\Pi^{ij}$ is the anisotropic pressure. Note that $\Pi^{ij}$ has no time-like components; it is symmetrical in its indices ($\Pi^{ij} = \Pi^{ji}$); and it is traceless ($\Pi^i_{\>\> i} = 0$). Thus we may write $\Pi_{ij}$ in terms of a projection operator:
\beq
\Pi_{ij} = \left[ \partial_i \partial_j - \frac{1}{3} \delta_{ij} \nabla^2 \right]  \Pi .
\label{Piprojection}
\eeq
As we will see below, $\Pi_{ij} \neq 0$ in models with at least one nonminimally coupled scalar field. The decomposition of $T_{\mu\nu}$ in Eq. (\ref{fluid}) is completely general: this form applies (with different values of $\rho$, $\delta \rho$, and so on) for simple models involving one minimally coupled field as well as for models with several nonminimally coupled fields. 

To background order, the Einstein equations of Eq. (\ref{Einstein}) yield the usual dynamical equations:
\beq
\begin{split}
3H^2 &= \frac{1}{M_{\rm pl}^2} \rho , \\
2 \dot{H} + 3 H^2 &= - \frac{1}{M_{\rm pl}^2} p .
\end{split}
\label{BackgroundEqs}
\eeq
The first-order perturbed Einstein equations yield
\beq
3H \left( \dot{\Psi} + H \Phi \right) - \frac{1}{a^2} \nabla^2 \Psi = - \frac{1}{2M_{\rm pl}^2} \delta \rho ,
\label{00}
\eeq
\beq
\dot{\Psi} + H \Phi = - \frac{1}{2M_{\rm pl}^2} \delta q ,
\label{0i}
\eeq
\beq
\ddot{\Psi} + 3 H \dot{\Psi} + H \dot{\Phi} + \left( 2 \dot{H} + 3 H^2 \right) \Phi = \frac{1}{2M_{\rm pl}^2} \left[ \delta p - \frac{2}{3} \nabla^2  \Pi \right] ,
\label{ij}
\eeq
and
\beq
\frac{1}{a^2} \partial^i \partial_j ( \Phi - \Psi ) = \frac{1}{M_{\rm pl}^2} \partial^i \partial_j  \Pi ,
\label{ineqj}
\eeq
where Eq. (\ref{00}) follows from the $00$ component of the Einstein field equations; Eq. (\ref{0i}) from the $0i$ component; Eq. (\ref{ij}) from the $i = j$ component; and Eq. (\ref{ineqj}) from the $i \neq j$ component.

Because of the Bianchi identity, the covariant derivative of the lefthand side of the Einstein field equation vanishes identically, $G^{\mu\nu}_{\>\>\>\>; \nu} = 0$. Thanks to Eq. (\ref{Rmn}), that implies energy-momentum conservation, $T^{\mu\nu}_{\>\>\>\>; \nu} = 0$. Upon calculating the Christoffel symbols to first order in the metric perturbations and keeping terms linear in the perturbations, we find
\beq
\begin{split}
T^{0\nu}_{\>\>\>\>; \nu} &= \left[ \dot{\rho} + 3 H (\rho + p) \right] (1 - 2 \Phi) \\
&\quad\quad\quad + \delta \dot{\rho} + 3 H (\delta \rho + \delta p) + \frac{1}{a^2} \nabla^2 \delta q - 3 (\rho + p) \dot{\Psi} = 0 .
\end{split}
\label{T0nunu}
\eeq
We consider the background and first-order terms to be separately conserved, which yields
\beq
\begin{split}
\dot{\rho} + 3 H (\rho + p) &= 0 , \\
\delta \dot{\rho} + 3 H (\delta \rho + \delta p ) &= - \frac{1}{a^2} \nabla^2 \delta q + 3 (\rho + p ) \dot{\Psi} .
\end{split}
\label{Energycons}
\eeq
Note that the anisotropic pressure, $\Pi_{ij}$, drops out of the conservation equations because it is traceless.

Comparing Eqs. (\ref{Tmn}) and (\ref{fluid}), we may identify the fluid components ($\rho$, $\delta \rho$, and so on) in terms of the matter-field content of our family of models. We expand the nonminimal coupling term as
\beq
f (\phi^I) = f (\varphi^I ) + \delta f (\varphi^I, \delta \phi^I ) + {\cal O} (\delta \phi^I \delta \phi^J ),
\label{expandf}
\eeq
where $f (\varphi^I) = f^{(0)}$. In what follows we will drop the superscript $(0)$ on $f$; it should be understood that $f$ written with no explicit argument refers to $f (\varphi^I)$. Then the $00$ component yields
\beq
\begin{split}
\rho &= \frac{M_{\rm pl}^2}{2f } \left[ \frac{1}{2} \delta_{IJ} \dot{\varphi}^I \dot{\varphi}^J + V - 6 H \dot{f} \right] , \\
\delta \rho &= \frac{M_{\rm pl}^2}{2f} \left[ \delta_{IJ} \left( \dot{\varphi}^I \delta \dot{\phi}^J - \dot{\varphi}^I \dot{\varphi}^J \Phi \right) + V_K \delta \phi^K  \right. \\
&\quad\quad\quad \quad \left. + 6 \dot{f} \left( \dot{\Psi} + 2 H \Phi \right) - 6 H \left( \delta \dot{f} + H \delta f \right) +  \frac{2}{a^2} \nabla^2 \delta f \right] .
\end{split}
\label{rhodeltarhoIJ}
\eeq
The $0i$ component becomes
\beq
\delta q = - \frac{M_{\rm pl}^2}{2f} \left[ \delta_{IJ} \dot{\varphi}^I \delta \phi^J + 2 \left(\delta \dot{f} - H \delta f - \dot{f}\Phi \right) \right] .
\label{deltaq}
\eeq
The diagonal terms within $T^i_{\>\>j}$ yield
\beq
\begin{split}
p &= \frac{M_{\rm pl}^2}{2f} \left[ \frac{1}{2} \delta_{IJ} \dot{\varphi}^I \dot{\varphi}^J - V + 2 \ddot{f} + 4 H \dot{f} \right] , \\
\delta p &= \frac{M_{\rm pl}^2}{2f} \left[ \delta_{IJ} \left( \dot{\varphi}^I \delta \dot{\phi}^J - \dot{\varphi}^I \dot{\varphi}^J \Phi \right) - V_K \delta \phi^K  - 4 \ddot{f} \Phi - 2 \dot{f} \left( \dot{\Psi} + 2 H \Phi \right) \right. \\
&\quad\quad\quad\quad \left. - \frac{2}{M_{\rm pl}^2} p \delta f + 2 \delta \ddot{f} + 4 H \delta \dot{f} -  \frac{2}{a^2} \nabla^2 \delta f \right] .
\end{split}
\label{pdeltap}
\eeq
The $i \neq j$ term, coming from the anisotropic pressure, becomes
\beq
\partial_i \partial_j (\Phi - \Psi ) = - \frac{1}{f} \partial_i \partial_j \delta f.
\label{PhiPsif}
\eeq
Thus in models with at least one nonminimally coupled scalar field, the two Bardeen potentials, $\Phi$ and $\Psi$, will differ. In the limit $f (\phi^I) \rightarrow M_{\rm pl}^2 / 2 = {\rm constant}$, Eqs. (\ref{rhodeltarhoIJ})-(\ref{pdeltap}) approach the corresponding expressions for the multifield, minimally coupled case (see Eqs. (72)-(74) of \cite{BTW}), and $\Phi \rightarrow \Psi$.

\section{Conservation of the Curvature Perturbation}

We may now consider the behavior of the curvature perturbation on uniform-density hypersurfaces, $\zeta$ \cite{BST}. In terms of the gauge-invariant Bardeen potential, $\Psi$, and working in longitudinal gauge, $\zeta$ may be written \cite{BTW}
\beq
\zeta = - \Psi - \frac{H}{\dot{\rho}} \delta \rho = - \Psi + \frac{\delta \rho}{3 (\rho + p)}, 
\label{zeta}
\eeq
where the second expression follows from using Eq. (\ref{Energycons}) for $\dot{\rho}$. Taking the time derivative we find
\beq
\begin{split}
\dot{\zeta} &= - \dot{\Psi} + \frac{\delta \dot{\rho}}{3 (\rho + p)} - \frac{\delta \rho}{3 (\rho + p)^2} \left( \dot{\rho} + \dot{p} \right) \\
&= - \frac{H \delta p}{ (\rho + p)} - \frac{1}{3 (\rho + p)} \frac{1}{a^2} \nabla^2 \delta q - \frac{\dot{p} \delta \rho}{3 (\rho + p)^2} ,
\end{split}
\label{dotzeta1}
\eeq
upon using the expression for $\delta \dot{\rho}$ in Eq. (\ref{Energycons}). The non-adiabatic pressure is defined as \cite{BTW}
\beq
\delta p_{\rm nad} \equiv \delta p - \frac{\dot{p}}{\dot{\rho}} \delta \rho ,
\label{pnad}
\eeq
with which we may rewrite Eq. (\ref{dotzeta1}) as
\beq
\dot{\zeta} = - \frac{H}{(\rho + p)} \delta p_{\rm nad} - \frac{1}{3 (\rho + p) }\frac{1}{a^2} \nabla^2 \delta q .
\label{dotzeta2}
\eeq
This expression for $\dot{\zeta}$ is completely general and model-independent; it holds for any (physical) case in which the total energy-momentum tensor is conserved.

The $\delta q$ term is related to the shear, which (as we will see in a moment) is always suppressed in the long-wavelength limit, $k \ll aH$. Thus any deviation of $\dot{\zeta}$ from zero on cosmologically interesting length scales must arise from the presence of non-adiabatic pressure, $\delta p_{\rm nad}$. In other words, models which produce significant entropy perturbations can have $\dot{\zeta} \neq 0$ even in the limit $k \ll aH$.

To evaluate the $\nabla^2 \delta q$ term in Eq. (\ref{dotzeta2}) we use Eqs. (\ref{00}) and (\ref{0i}), relating $\delta q$ and $\delta \rho$ to $\Psi$ and $\Phi$. We also perform a Fourier transform, such that $\nabla^2 F = - k^2 F$ for any function $F(x^i)$, where $k$ is the comoving wavenumber. Then we find
\beq
\begin{split}
\frac{1}{3 (\rho + p)} \frac{k^2}{a^2} \delta q &= - \frac{2 M_{\rm pl}^2}{3 (\rho + p)} \frac{k^2}{a^2} \left(\dot{\Psi} + H \Phi \right) \\
&= \frac{1}{3} H \left( \frac{k}{aH} \right)^2 \left[ \zeta + \Psi \left( 1 + \frac{2 \rho}{9 (\rho + p)} \left( \frac{k}{aH} \right)^2 \right) \right] \equiv - \Sigma ,
\end{split}
\label{Sigma}
\eeq
where $\Sigma$ is the scalar shear along comoving worldlines \cite{LythWands,BTW}, and we have used the definition of $\zeta$ in Eq. (\ref{zeta}) as well as the background relation between $H^2$ and $\rho$ of Eq. (\ref{BackgroundEqs}). Clearly the shear will remain negligible on the relevant length scales following Hubble crossing, with $k \ll aH$, so long as $\zeta$ and $\Psi$ remain finite. This result is model-independent, and holds for any conserved energy-momentum tensor. Returning to Eq. (\ref{dotzeta2}), we thus see that any deviations of $\dot{\zeta}$ from zero in the limit $k \ll aH$ will arise from entropy perturbations, $\delta p_{\rm nad}$.

Consider first the case of a single scalar field with minimal coupling. Applying Eqs. (\ref{eomvarphideltaphi}), (\ref{BackgroundEqs}), and (\ref{rhodeltarhoIJ})-(\ref{pdeltap}) in the case $N = 1$ and $f (\varphi^I) = M_{\rm pl}^2 / 2$, $\delta f = 0$, we find
\beq
\frac{\dot{p}}{\dot{\rho}} = 1 + \frac{2 V_\phi}{3 H \dot{\varphi}} ,
\label{dotpdotrhovanilla}
\eeq
and thus, from Eq. (\ref{pnad}),
\beq
\begin{split}
\delta p_{\rm nad} &= - 2 V_\phi \delta \phi - \frac{2V_\phi}{3H\dot{\varphi}} \delta \rho = - \frac{2V_\phi}{3H\dot{\varphi}} \left( \delta \rho + 3 H \dot{\varphi} \delta \phi \right) \\
&= - \frac{2V_\phi}{3H\dot{\varphi}} \delta \rho_m ,
\end{split}
\label{pnadvanilla1}
\eeq
where $\delta \rho_m$ is the gauge-invariant comoving density perturbation \cite{BTW}, defined as
\beq
\delta \rho_m = \delta \rho - 3 H \delta q .
\label{deltarhom}
\eeq
In the single-field, minimally coupled case, we thus find that $\delta p_{\rm nad} \propto \delta \rho_m$. We may further evaluate $\delta \rho_m$ by combining Eqs. (\ref{00}) and (\ref{0i}), which yields
\beq
\frac{k^2}{a^2} \Psi = - \frac{1}{2M_{\rm pl}^2} (\delta \rho - 3 H \delta q) = - \frac{1}{2M_{\rm pl}^2} \delta \rho_m .
\label{gradPsirhom}
\eeq
This expression, relating $\delta \rho_m$ to the spatial derivative of $\Psi$, is completely general and holds for any conserved energy-momentum tensor; the proportionality of $\delta p_{\rm nad}$ and $\delta \rho_m$ in Eq. (\ref{pnadvanilla1}) is model specific. In the single-field, minimally coupled case, we therefore find
\beq
\delta p_{\rm nad} = - \frac{2V_\phi}{3 H\dot{\varphi}} \delta \rho_m = \frac{4 V_\phi \rho}{9 H \dot{\varphi}} \left( \frac{k}{aH} \right)^2 \Psi .
\label{pnadvanilla2}
\eeq
In the simple case of a single minimally coupled field, we find the well-known result \cite{WMLL,BST,MFB,BTW,LythWands} that both $\delta p_{\rm nad}$ and $\Sigma$ are suppressed by factors of $(k / aH)^2$. From Eq. (\ref{dotzeta2}), we therefore see that in single-fleld, minimally coupled models, the curvature perturbation remains conserved, $\dot{\zeta} \simeq 0$, on cosmologically relevant length scales with $k \ll aH$. In such models, curvature perturbation modes that are amplified during inflation and cross outside the Hubble radius will remain effectively frozen until crossing back inside the Hubble radius at the time of CMB last scattering.

We may now use the formalism of Section II to see how this result generalizes to the case of multiple scalar fields, each with nonminimal couplings. From Eq. (\ref{rhodeltarhoIJ}), we find
\beq
\dot{\rho} = \frac{\dot{f}}{f} \rho - \frac{3H M_{\rm pl}^2}{2f} \left[ \delta_{IJ} \dot{\varphi}^I \dot{\varphi}^J + 2 \ddot{f} \right] ,
\label{dotrhomulti}
\eeq
upon using Eq. (\ref{eomvarphideltaphi}) for $\ddot{\varphi}^I$, Eq. (\ref{RdeltaR}) for $R$, Eq. (\ref{BackgroundEqs}) to relate $H^2$ and $\rho$, and the relation $\dot{f} = f_I \dot{\varphi}^I$. Proceeding similarly, from Eq. (\ref{pdeltap}) we find
\beq
\dot{p} = \dot{\rho} - \frac{3}{2} \frac{\dot{f}}{f} (\rho + p) - 2 \frac{\dot{f}}{f} p + \frac{M_{\rm pl}^2}{2f} \left[ - 2 V_I \dot{\varphi}^I + 2 \dddot{f} + 10 H \ddot{f} \right] .
\label{dotpmulti1}
\eeq
From Eqs. (\ref{BackgroundEqs}) and (\ref{Energycons}) we may write
\beq
(\rho + p ) = -\frac{1}{3H} \dot{\rho} = - 2 M_{\rm pl}^2 \dot{H} .
\label{rhoplusp}
\eeq
Combining, we find
\beq
\frac{\dot{p}}{\dot{\rho}} = 1 + \left( \frac{M_{\rm pl}^2}{2f} \right) \frac{2V_I \dot{\varphi}^I}{3H (\rho + p)} + {\cal S} ,
\label{dotpdotrho}
\eeq
where we have defined
\beq
{\cal S} \equiv \frac{1}{2H} \frac{\dot{f}}{f} \left[ 1 + \frac{4p}{3(\rho + p)} \right] + \frac{1}{6 H \dot{H}} \left[ \frac{\dddot{f} }{f} + 5 H \frac{\ddot{f}}{f} \right] ,
\label{calS}
\eeq
and we have again used $f = f (\varphi^I) = f^{(0)}$ for the nonminimal coupling function evaluated in terms of the background fields, $\varphi^I$. Similar calculation yields
\beq
\delta p - \delta \rho = \left( \frac{M_{\rm pl}^2}{2f} \right) \left[ - 2 V_I \delta \phi^I - {\cal F} + \delta {\cal F} \right],
\label{deltapdeltarho}
\eeq
where
\beq
\begin{split}
{\cal F} &\equiv 4 \ddot{f} \Phi + 2 \dot{f} \dot{\Phi} + 10 \dot{f} \left( \dot{\Psi} + 2 H \Phi \right) , \\
\delta {\cal F} &\equiv \frac{2}{M_{\rm pl}^2} (\rho - p) \delta f + 2 \delta \ddot{f} + 10 H \delta \dot{f} + 4 \frac{k^2}{a^2} \delta f .
\end{split}
\label{calF}
\eeq
Combining Eqs. (\ref{dotpdotrho})-(\ref{calF}) and using Eq. (\ref{deltaq}) for $\delta q$, we thus find for Eq. (\ref{pnad}),
\beq
\delta p_{\rm nad} = - \left( \frac{M_{\rm pl}^2}{2f} \right) \left[ \frac{2V_I \dot{\varphi}^I}{3H (\rho + p)} \delta \rho_m + 2 V_I \Delta^I + {\cal F} - \delta {\cal F} - \left( \frac{2f}{M_{\rm pl}^2} \right) {\cal S} \delta \rho \right] ,
\label{pnad2}
\eeq
where we have defined
\beq
\Delta^I \equiv \delta \phi^I + \frac{\delta q}{(\rho + p)} \dot{\varphi}^I .
\label{DeltaI}
\eeq
In the minimally coupled case, $f (\varphi^I) \rightarrow M_{\rm pl}^2 / 2 = {\rm constant}$ and $\delta f \rightarrow 0$, and thus $({\cal S}, {\cal F}, \delta {\cal F}) \rightarrow 0$.

We may quickly confirm the well-known result that multifield models, each with canonical kinetic terms and minimal coupling, generically produce entropy perturbations that need not be suppressed in the limit $k \ll aH$ \cite{GBW,BFKM,GWBM,RFB2002,WandsReview,BTW}. For simplicity, consider a two-field model with fields $\phi$ and $\chi$. Then Eq. (\ref{pnad2}) becomes
\beq
\delta p_{\rm nad} = - \frac{(2V_\phi \dot{\phi} + 2 V_\chi \dot{\chi} )}{3H (\dot{\phi}^2 + \dot{\chi}^2)} \delta \rho_m - \frac{2 \dot{\phi} \dot{\chi}}{(\dot{\phi}^2 + \dot{\chi}^2)} \left( \dot{\chi} V_\phi - \dot{\phi} V_\chi \right) \left[ \frac{\delta \phi}{\dot{\phi}} - \frac{\delta \chi}{\dot{\chi}} \right] .
\label{pnadmultiMC}
\eeq
As usual, the term proportional to $\delta \rho_m$ will remain suppressed in the long-wavelength limit, thanks to Eq. (\ref{gradPsirhom}), but the second term need not be negligible even for $k \ll aH$. Multiple interacting fields will generically produce entropy perturbations which can in turn amplify the curvature perturbation.

From Eq. (\ref{pnad2}) it is also clear that nonminimal couplings will generate a second, distinct source of entropy perturbations, which likewise need not be suppressed in the limit $k \ll aH$. To distinguish this second source of entropy perturbations from those that are generically produced in the multifield case, consider a model with a single nonminimally coupled scalar field. Upon using Eqs. (\ref{rhodeltarhoIJ})-(\ref{pdeltap}) in the case $N = 1$, $(f, \delta f) \neq {\rm constant}$, Eq. (\ref{pnad2}) becomes
\beq
\delta p_{\rm nad} = -\left( \frac{M_{\rm pl}^2}{2f} \right) \left[ \frac{2V_\phi \dot{\varphi}}{3H (\rho + p)} \delta \rho_m + 2V_\phi \Delta (\phi) + {\cal F} - \delta {\cal F} + \left(\frac{2f}{M_{\rm pl}^2} \right) {\cal S} \delta \rho \right] ,
\label{pnadNMC}
\eeq
where ${\cal S}$ is defined in Eq. (\ref{calS}), ${\cal F}$ and $\delta {\cal F}$ are defined in Eq. (\ref{calF}), and, in the single-field case, Eq. (\ref{DeltaI}) reduces to
\beq
\Delta (\phi) = \frac{2}{[ \dot{\varphi}^2 + 2 \ddot{f} - 2 H \dot{f}]} \left[\dot{\varphi} \left( \delta \dot{f} - H \delta f - \dot{f} \Phi \right) - \delta \phi \left( \ddot{f} - H \dot{f} \right) \right]
\label{Deltaphi}
\eeq
From Eqs. (\ref{gradPsirhom}) and (\ref{pnadNMC}) we again see that the first term in $\delta p_{\rm nad}$, proportional to $\delta \rho_m$, will remain suppressed for $k \ll aH$, whereas all of the remaining terms --- which arise solely from the nonminimal coupling --- can source entropy perturbations even in the long-wavelength limit.

The behavior of the curvature perturbation, $\zeta$, thus depends upon the entropy perturbations, $\delta p_{\rm nad}$. The entropy perturbations, in turn, depend on several terms involving the nonminimal couplings: $\Delta^I$, ${\cal F}$, $\delta {\cal F}$, and ${\cal S} \delta \rho$. From Eqs. (\ref{rhodeltarhoIJ}), (\ref{deltaq}), (\ref{calS}), (\ref{calF}), and (\ref{DeltaI}), we see that $\Delta^I$, ${\cal F}$, and ${\cal S} \delta \rho$ are each proportional to the field fluctuations, $\delta \phi^I$, while the ${\cal F}$ term depends on the time-dependence of the nonminimal coupling function, $f (\varphi^I)$. During inflation, we expect all of these contributions to $\delta p_{\rm nad}$ to remain negligible, at least in the limit $k \ll aH$. The field fluctuations in many models will remain close to the usual value in de Sitter spacetime, $\langle \delta \phi^I \rangle \sim H / 2\pi$. Likewise, the background fields, $\varphi^I$, should vary slowly during the slow-roll phase of inflation, so that $\dot{f}$, $\ddot{f}$, and similar terms should remain relatively unimportant. 

As inflation ends and preheating begins, however, both of these conditions can change dramatically. The inflaton, $\varphi$, will begin to oscillate rapidly, and hence $\dot{f}$, $\ddot{f}$, and similar terms need not remain small. Furthermore, under certain conditions, the field fluctuations, $\delta \phi^I_k$, can become resonantly amplified, growing exponentially even for super-Hubble modes with $k \ll aH$. (For reviews of preheating, see \cite{BTW,ABCM}.) Thus during preheating, the entropy perturbations arising from nonminimal couplings can grow rapidly. This growth, in turn, can amplify the entropy perturbations, $\delta p_{\rm nad}$, and drive $\dot{\zeta} \neq 0$. 

\section{Conclusions}

Models with multiple nonminimally coupled scalar fields generically produce two distinct sources of entropy perturbations. When analyzed in the Jordan frame, these two sources are easily distinguished. One set arises strictly from the interactions of multiple fields (and persists even when all fields are minimally coupled), and a second set arises strictly from nonminimal couplings (even in the case of a single field). The term $\Delta^I$ of Eq. (\ref{DeltaI}) contributes to both sources of entropy perturbations, because the quantities $\rho$, $p$, and $\delta q$ acquire terms dependent on the nonminimal couplings, $f (\phi^I)$. The additional new terms that we have identified --- ${\cal S}$, ${\cal F}$, and $\delta {\cal F}$ in Eqs. (\ref{calS}) and (\ref{calF}) --- each contribute entropy perturbations due to the presence of nonminimal couplings. All of these terms arise for fields with canonical kinetic terms in the action; and none of the new sources of entropy perturbations will necessarily be suppressed in the limit $k \ll aH$.

The various sources of entropy perturbations may be understood intuitively. Consider pouring multiple fluids into a container. Mixing the fluids will naturally produce entropy; think of how difficult it would be to un-mix the fluids after they have been poured in. That source of entropy corresponds to the entropy perturbations present in any multifield model. Next imagine that the walls of the container were allowed to wobble in response to the sloshing of the fluids. The walls' vibrations would further increase the number of allowable states toward which the system could evolve. That effect is akin to the nonminimal couplings, which introduce an added degree of freedom to the way spacetime responds to the presence of matter. (In effect, the nonminimal couplings make the local strength of gravity depend on space and time, $G \propto 1/ f(\phi^I)$.) Hence the nonminimal couplings contribute a separate source of entropy to the system.

In the single-field case, one may always perform a conformal transformation and field rescaling so that the dynamics (in terms of the rescaled field) appear identical to the single-field minimally coupled case, akin to Eq. (\ref{pnadvanilla1}) \cite{Hwang,NewTwist,Christopherson}. But in the multifield case, no combination of conformal transformation and field rescalings exists that could make all fields appear to have both minimal couplings to $R$ and canonical kinetic terms in the action \cite{DKconftrans}. The new source of entropy perturbations identified here, which arises from nonminimal couplings in the Jordan frame, would appear in the transformed frame to arise from fields' noncanonical kinetic terms \cite{RFB2002,Spinflation,Langlois}. One way or another, in the multifield case, there exist two separate sources of entropy perturbations even in the long-wavelength limit.

\acknowledgements{It is a pleasure to thank Bruce Bassett for helpful discussions. ATT also acknowledges Prof. Edmund Bertschinger and Prof. Edward Farhi for facilitating her visit at MIT's Center for Theoretical Physics. This work was supported in part by the U.S. Department of Energy (DoE) under contract No. DE-FG02-05ER41360.}


\begin{thebibliography}{999}

\bibitem{Komatsu} E. Komatsu et al. [WMAP collaboration], arXiv:1001.4538 [astro-ph.CO].

\bibitem{GuthKaiser} A. H. Guth and D. I. Kaiser, Science {\bf 307}, 884 (2005) [arXiv:astro-ph/0502328].

\bibitem{MFB} V. F. Mukhanov, H. A. Feldman, and R. H. Brandenberger, Phys. Rept. {\bf 215}, 203 (1992).

\bibitem{RHB2003} R. H. Brandenberger, Lect. Notes. Phys. {\bf 646}, 127 (2004) [arXiv:hep-th/0306071].

\bibitem{BTW} B. A. Bassett, S. Tsujikawa, and D. Wands, Rev. Mod. Phys. {\bf 78}, 537 (2005) [arXiv:astro-ph/0507632].

\bibitem{LiddleLythBook} D. H. Lyth and A. R. Liddle, {\it The Primordial Density Perturbation: Cosmology, Inflation, and the Origin of Structure} (New York: Cambridge University Press, 2009).

\bibitem{MalikWands} K. A. Malik and D. Wands, Phys. Rept. {\bf 475}, 1 (2009) [arXiv:0908.4944 [astro-ph]].

\bibitem{GWBM} C. Gordon, D. Wands, B. A. Bassett, and R. Maartens, Phys. Rev. {\bf D63}, 023506 (2001) [arXiv:astro-ph/0009131].

\bibitem{BassettTsujikawa} S. Tsujikawa and B. A. Bassett, Phys. Lett. {\bf B536}, 9 (2002) [arXiv:astro-ph/0204031].

\bibitem{WandsReview} D. Wands, Lect. Notes. Phys. {\bf 738}, 275 (2008) [arXiv:astro-ph/0702187].

\bibitem{RFB2002} F. Di Marco, F. Finelli, and R. H. Brandenberger, Phys. Rev. {\bf D67}, 063512 (2003) [arXiv:astro-ph/0211276].

\bibitem{Spinflation} D. A. Easson, R. Gregory, D. F. Mota, G. Tasinato, and I. Zavala, JCAP {\bf 0802}, 010 (2008) [arXiv:0709.2666 [hep-th]].

\bibitem{Langlois} D. Langlois and S. Renaux-Petel, JCAP {\bf 0804}, 017 (2008) [arXiv:0801.1085 [hep-th]].

\bibitem{Fujii} Y. Fujii and K. Maeda, {\it The Scalar-Tensor Theory of Gravitation} (New York: Cambridge University Press, 2003).

\bibitem{Faraoni2004} V. Faraoni, {\it Cosmology in Scalar-Tensor Gravity} (Boston: Kluwer, 2004).

\bibitem{BirrellDavies} N. D. Birrell and P. C. W. Davies, {\it Quantum Fields in Curved Space} (New York: Cambridge University Press, 1982).

\bibitem{Buchbinder} I. L. Buchbinder, S. D. Odintsov, and I. L. Shapiro, {\it Effective Action in Quantum Gravity} (New York: Taylor and Francis, 1992).

\bibitem{DKconftrans} D. I. Kaiser, Phys. Rev. {\bf D81}, 084044 (2010) [arXiv:1003.1159 [gr-qc]].

\bibitem{Hwang} J. C. Hwang, Class. Quant. Grav. {\bf 7}, 1613 (1990); J. C. Hwang, Phys. Rev. {\bf D42}, 2601 (1990); J. C. Hwang, Class. Quant. Grav. {\bf 8}, 195 (1991); J. C. Hwang, Astrophys. J. {\bf 375}, 443 (1991).

\bibitem{HiraiMaeda} T. Hirai and K.-I. Maeda, Astrophys. J. {\bf 431}, 6 (1994) [arXiv:astro-ph/9404023].

\bibitem{NMCPerts} See, e.g., A. H. Guth and B. Jain, Phys. Rev. {\bf D45}, 426 (1992); D. I. Kaiser, Phys. Lett. {\bf B340}, 23 (1994) [arXiv:astro-ph/9405029]; D. I. Kaiser, Phys. Rev. {\bf D52}, 4295 (1995) [arXiv:astro-ph/9408044]; A. A. Starobinsky and J. Yokoyama, arXiv:gr-qc/9502002; S. Tsujikawa and H. Yajima, Phys. Rev. {\bf D62}, 123512 (2000) [arXiv:hep-ph/0007351]; A. A. Starobinsky, S. Tsujikawa, and J. Yokoyama, Nucl. Phys. {\bf B610}, 383 (2001) [arXiv:astro-ph/0107555]; M. Yu. Khlopov and S. G. Rubin, {\it Cosmological Pattern of Microphysics in the Inflationary Universe} (Kluwer Academic Publishers, Dordrecht, 2004); S. Tsujikawa and B. Gumjudpai, Phys. Rev. {\bf D69}, 123523 (2004) [arXiv:astro-ph/0402185]; Z. Lalak, D. Langlois, S. Pokorski, and K. Turzynski, JCAP {\bf 0707}, 014 (2007) [arXiv:0704.0212 [hep-th]]; S. Tsujikawa, K. Uddin, and R. Tavakol, Phys. Rev. {\bf D77}, 043007 (2008) [arXiv:0712.0082]; D. Langlois, S. Renaux-Petel, D. A. Steer, and T. Tanaka, Phys. Rev. Lett. {\bf 101}, 061301 (2008) [arXiv:0804.3139 [hep-th]]; D. Langlois, S. Renaux-Petel, D. A. Steer, and T. Tanaka, Phys. Rev. {\bf D78}, 063523 (2008) [arXiv:0806.0336 [hep-th]]; S. Renaux-Petel and G. Tasinato, JCAP {\bf 0901}, 012 (2009) [arXiv:0810.2405 [hep-th]]; X. Ji and T. Wang, arXiv:0903.0379 [hep-th]; X. Gao, arXiv:0908.4035 [hep-th]; A. De Felice and S. Tsujikawa, arXiv:1002.4928 [gr-qc]; C.-J. Feng, X.-Z. Li, and E. N. Saridakis, arXiv:1004.1874 [astro-ph.CO].

\bibitem{NewTwist} S. Tsujikawa and B. A. Bassett, Phys. Rev. {\bf D62}, 043510 (2000) [arXiv:hep-ph/0003068]. 

\bibitem{Christopherson} A. J. Christopherson and K. A. Malik, Phys. Lett. {\bf B675}, 159 (2009) [arXiv:0809.3518 [astro-ph]].

\bibitem{Faraoni2001} V. Faraoni, Int. J. Theor. Phys. {\bf 40}, 2259 (2001) [arXiv:hep-th/0009053].

\bibitem{BurgessHertzberg} C. P. Burgess, H. M. Lee, and M. Trott, JHEP {\bf 0909}, 103 (2009) [arXiv:0902.4465 [hep-ph]]; C. P. Burgess, H. M. Lee, and M. Trott, arXiv:1002.2730 [hep-ph]; M. P. Hertzberg, arXiv:1002.2995 [hep-ph].

\bibitem{LythRiottoMazumdar} D. H. Lyth and A. Riotto, Phys. Rept. {\bf 314}, 1 (1999) [arXiv:hep-ph/9807278]; A. Mazumdar and J. Rocher, arXiv:1001.0993 [hep-ph].

\bibitem{ShapBez} F. L. Bezrukov and M. E. Shaposhnikov, Phys. Lett. {\bf B659}, 703 (2008) [arXiv:0710.3755 [hep-th].

\bibitem{WeinbergQFT} S. Weinberg, {\it The Quantum Theory of Fields}, vol. 2, {\it Modern Applications} (New York: Cambridge University Press, 1996).

\bibitem{BST} J. M. Bardeen, P. J. Steinhardt, and M. S. Turner, Phys. Rev. {\bf D28}, 679 (1983).

\bibitem{LythWands} D. H. Lyth and D. Wands, Phys. Rev. {\bf D68}, 103515 (2003) [arXiv:astro-ph/0306498].

\bibitem{WMLL} D. Wands, K. A. Malik, D. H. Lyth, and A. R. Liddle, Phys. Rev. {\bf D62}, 043527 (2000) [arXiv:astro-ph/0003278].

\bibitem{GBW} J. Garcia-Bellido and D. Wands, Phys. Rev. {\bf D52}, 6739 (1995) [arXiv:gr-qc/9506050].

\bibitem{BFKM} B. A. Bassett, F. Tamburini, D. I. Kaiser, and R. Maartens, Nucl. Phys. {\bf B561}, 188 (1999) [arXiv:hep-ph/9901319]; B. A. Bassett, C. Gordon, R. Maartens, and D. I. Kaiser, Phys. Rev. {\bf D61}, 061302 (2000) [arXiv:hep-ph/9909482].

\bibitem{ABCM} R. Allahverdi, R. Brandenberger, F.-Y. Cyr-Racine, and A. Mazumdar, arXiv:1001.2600 [hep-th].

\end{thebibliography}
\end{document}